\documentclass[12pt,a4paper]{article}
\usepackage[english]{babel}
\usepackage[latin1]{inputenc}
\usepackage[T1]{fontenc}
\usepackage{amsmath}
\usepackage{amsfonts}
\usepackage{amssymb}
\usepackage{mathtools}
\usepackage{braket}
\usepackage{booktabs}
\usepackage{dsfont}
\usepackage{lmodern,bm}
\usepackage{slashed}
\usepackage{float}
\usepackage{caption}
\usepackage{graphicx}
\usepackage{hyperref}
\begin{document}
\title{On the fermionic van der Waals and Casimir-Polder interactions} 
\author{C.~D.~Fosco and G.~Hansen\\
{\normalsize\it Centro At\'omico Bariloche and Instituto Balseiro}\\
{\normalsize\it Comisi\'on Nacional de Energ\'{\i}a At\'omica}\\
{\normalsize\it R8402AGP S.\ C.\ de Bariloche, Argentina.}}

\maketitle
\begin{abstract}
We formulate fermionic versions, for any number of spatial dimensions,
of the van der Waals and Casimir-Polder interactions, and study their properties.  
In both cases, the systems we introduce contain localized `atoms': two-level 
systems, coupled to a vacuum Dirac field. This Dirac field plays here a role 
akin to the electromagnetic field in the van der Waals case. 

\noindent In this context, bag-model conditions for the Dirac field serve as 
the analog of the 'mirror' in the Casimir-Polder effect.
We found that, in this case, the resulting interaction is repulsive.

\end{abstract}

\section{Introduction}\label{sec:intro}
Some distinctive effects due to the quantum vacuum~\cite{milonni1994quantum} 
have become the subject of intensive research, partly due to the fact that those 
effects may become relevant, even dominant, at the increasingly accessible 
nanoscopic scales~\cite{lamoreaux2005casimir}. 
Among them is the van der Waals interaction, by which
two electrically neutral microscopic objects may interact, as a consequence
of the quantum fluctuations of their respective dipole operators. They 
fluctuate with zero average, but the electromagnetic (EM) field coupling 
leads to a non-vanishing interaction energy, which depends on the quantum 
average of  the {\em  square\/} of each dipole moment. A closely related 
phenomenon is 
the  Casimir-Polder effect, by which one of those microscopic objects
facing a reflecting wall experiences an attractive force. This force may
be interpreted as due to the presence of an image atom, which also
fluctuates, albeit in coordination with the real one.

The Casimir effect is sometimes explained as the result of the effect of 
van der Waals interactions between the constituents of the media which
compose the `mirrors'. It is, however, clear, that those interactions do
{\em not\/} obey a superposition principle~\cite{CcapaTtira:2009ku}, and as a 
consequence the total effect cannot  be obtained, except in very special 
circumstances, as a sum of the pairwise  interactions. 

Quantum interactions involving fermionic fields are less explored compared to 
their electromagnetic counterparts. This work aims to bridge this gap by 
formulating and analyzing fermionic analogs of van der Waals and Casimir-Polder 
interactions. 
In the  fermionic Casimir effect, a vacuum Dirac field is in the presence of 
surfaces where the normal component of the current vanishes. 

In this paper we study that kind  of phenomenon, namely, how
localized objects can be coupled to a fermionic field, leading to a quantum 
interaction similar to its known EM counterpart.
More explicitly, we consider either one or two two-level systems, which 
represent localized degrees of freedom, coupled to a massless Dirac field.  In 
the case of just one two-level 
system, it will be facing a wall which imposes bag model boundary
conditions on the fermionic field. 

Realizations of this kind of system might arise, for instance, in
graphene, where a Dirac field in $2+1$ dimensions appears as an effective
description. Here, different kinds of localized defects, affecting the
Dirac field, naturally appear~\cite{PhysRevB.108.054101}.

In a rather different context, interesting phenomena  on the spin-dependent 
interactions between fermions, due to the exchange of virtual particles, 
have been studied in~\cite{Fichet:2017bng,Costantino:2019ixl}, and 
in~\cite{Costantino:2020bei}, an example where virtual neutrinos induce 
Casimir and Casimir-Polder like effects has been presented.

In this paper, we carry out a study of this kind of interaction in different 
numbers of dimensions, showing that the quantum vacuum fluctuations lead to 
an effective interaction energy which becomes entirely analogous, albeit in 
a fermionic realm, to what happens in the van der Waals interaction and the 
Casimir-Polder effect.

This paper is organized as follows: in Sect.~\ref{sec:the_system}, we 
introduce the elements of the system considered in this paper. Then in 
Sect.~\ref{sec:effective_action}, we evaluate the effective action 
corresponding to two atoms, to determine the resulting interaction energy, 
result which is presented in Sect.~\ref{sec:vanderwaals}. 
Analogous results, but for the interaction between an atom and a wall are 
presented in Sect.~\ref{sec:casimir-polder}. 
Finally, in Sect.~\ref{sec:conclusions}, we present the conclusions.

\section{The system}\label{sec:the_system}
Throughout this paper, we shall consider either one or two localized
two-level systems, at fixed positions, coupled to a vacuum Dirac field.
Each two-level system will be described by a complex multi-component
Grassmann field: $\theta = (\theta_a)$, with $a =1, \ldots, n$, and
$\bar{\theta}=\theta^\dagger$, in such a way that its free Euclidean
action is given by:
\begin{equation}
    \label{eq:S0tl}
    \mathcal{S}_0^a(\theta , \bar{\theta})
    \;=\;
    \int d\tau \, \bar{\theta}(\tau) \, (\partial_\tau + \Omega ) \,
    \theta(\tau) \;,
\end{equation}
where $\tau$ denotes the imaginary time and $\Omega$ the energy of the
excited state, relative to the ground state energy. The number of 
components of $\theta$ is assumed to match that of a Dirac field (in the 
irreducible representation of the
respective Clifford algebra) in $d+1$ space-time dimensions, whose 
free action is assumed to be:
\begin{equation}
    \label{eq:S0_D}
    \mathcal{S}_0^D(\psi,\bar{\psi})
    \;=\;
    \int d^{d+1} x \, \bar{\psi}(x) \, \slashed{\partial} \, \psi(x) \, ,
\end{equation}
with $\slashed{\partial} \equiv \gamma_\mu \partial_\mu$ denoting the
Dirac operator in $d+1$ dimensions. Dirac $\gamma$-matrices are Hermitian
and assumed satisfy the anti-commutation relations:
\begin{equation}
    \{\gamma_\mu, \gamma_\nu\}
    = \gamma_\mu \gamma_\nu + \gamma_\nu
    \gamma_\mu
    = 2\,\delta_{\mu\nu} \,,
\end{equation}
where $\mu, \nu = 0, 1, \ldots, d$. They are chosen according to the
conventions used in \cite{book:ZinnJustinQFT2021}, but it should be
noted that the results presented in this work are independent of the
specific representation of the $\gamma$-matrices.

We use natural units, whereby $\hbar$ and the velocity corresponding to the 
dispersion relation for the fermions, $v_F$, are both equal to $1$.  
Space-time coordinates are denoted by
$x_\mu$, and we shall occasionally use $\tau = x_0$ and $\mathbf{x}$ for
the spatial coordinates.
The metric is Euclidean: $g_{\mu\nu} = \text{diag}(1, \ldots, 1)$, 
and no distinctions are meant  between upper and lower indices, which due 
their positions to notational convenience.
Einstein summation convention for repeated indices in monomial
expressions is also assumed, unless explicitly stated otherwise.

The total action $\mathcal{S}$, in the case of two
atoms, at the positions ${\mathbf r}^{(1)}$ and
$\mathbf{r}^{(2)}$
is then introduced, with the following form:
\begin{align}
    \mathcal{S}
    &\;=\;
    \mathcal{S}_0^a(\theta^{(1)},\bar{\theta}^{(1)})
    \,+\, \mathcal{S}_{0}^{a}(\theta^{(2)},\bar{\theta}^{(2)})
    \,+\, \mathcal{S}^D(\psi,\bar{\psi}; \theta, \bar\theta ;
    \,\mathbf{r}^{(1)},\mathbf{r}^{(2)}) \,,
\end{align}
where $\theta = (\theta^{(1)},\theta^{(2)})$ denotes the Grassmann degrees
of freedom sitting on two spatial positions, and $\mathcal{S}^D$ is the
Dirac field action,  now including  coupling to the two-level systems:
\begin{align}
    \mathcal{S}^D(\psi,\bar{\psi}; &\,\theta, \bar\theta ; 
    \,\mathbf{r}^{(1)},\mathbf{r}^{(2)})
    \;=\;
   \mathcal{S}_{0}^{D}(\psi,\bar{\psi}) \nonumber\\
    &\;+\;
    \lambda \, \sum_{\alpha=1}^2 \,
    \int d\tau \, \big( \,\bar{\psi}(\tau,\mathbf{r}^{(\alpha)}) \,
    \theta^{(\alpha)}(\tau)
    \;+\; \bar{\theta}^{(\alpha)}(\tau) \, \psi(\tau,\mathbf{r}^{(\alpha)}) 
    \, \big) \, ,
\end{align}
where the constant $\lambda$ sets the strength of the coupling between
the atoms and the field. 

It is worth noting here some features of the proposed system, in the
context of the phenomena we want to describe: we first note that the
effect of the  atoms, when integrating out their degrees of freedom, would be to 
introduce a kind of frequency  dependent mass term, for the Dirac field. 
That mass goes like the square of the constant  $\lambda$, and is localized on the 
position (center of mass) of the atom. 
We recall that the bag model conditions may be obtained by introducing mass 
terms localized on the boundaries~\cite{Fosco:2008vn}, thus, the model we consider 
here appears to be an acceptable microscopic version of the EM
Casimir effect. 

\section{Effective action}\label{sec:effective_action}
We introduce the effective action $\Gamma_{\text{eff}}$ which results from 
the functional integration of all the degrees of freedom in the system:
\begin{align}
    e^{-\Gamma_{\text{eff}}(\mathbf{r}^{(1)},\mathbf{r}^{(2)})}
    &\;\equiv\;
    \frac{1}{\mathcal{N}} \, \mathcal{Z}
    (\mathbf{r}^{(1)},\mathbf{r}^{(2)}) 
    \,,  \nonumber \\
    \mathcal{Z}(\mathbf{r}^{(1)},\mathbf{r}^{(2)}) 
    &\;=\;
    \int \mathcal{D}\theta \mathcal{D}\bar{\theta} \mathcal{D}\psi
	\mathcal{D}\bar{\psi} \;\;
	e^{-\mathcal{S}
    (\theta,\bar{\theta},\psi,\bar{\psi}\,;\,\mathbf{r}^{(1)},
    \mathbf{r}^{(2)})} \; .
\end{align}
Since the atoms are static, this would be proportional to the vacuum energy 
and the total evolution time. 

The interaction energy of the system, subtracting the self-energy
contributions, is obtained by a careful choice of the  normalization
constant $\mathcal{N}$, namely, such that the interaction energy vanishes
in the limit of infinite distance between the atoms:
\begin{equation}
    \mathcal{N}
    \;\equiv\;
    \big[ \, \mathcal{Z}\big(\mathbf{r}^{(1)},\mathbf{r}^{(2)}\big) \, 
    \big]_{|\mathbf{r}^{(1)} - \mathbf{r}^{(2)}| \to \infty} \, .
\end{equation}
Note that with this in mind, we may discard terms that do not depend on the 
positions $\mathbf{r}^{(1)}$ and $\mathbf{r}^{(2)}$ in the intermediate 
calculations of the effective action.

Functionally integrating the Dirac field leads us to an
intermediate effective action, denoted by $\mathcal{S}_{\text{eff}}$
\begin{align}\label{eq:SeffAA}
    e^{-\mathcal{S}_{\text{eff}}(\theta, \bar{\theta}; \mathbf{r}^{(1)},
    \mathbf{r}^{(2)})} 
    &\;\equiv\; 
    \int \mathcal{D}\psi \mathcal{D}\bar{\psi} \, 
    e^{-\mathcal{S}(\theta, \bar{\theta}; \psi, \bar{\psi})} \nonumber\\
        \mathcal{S}_{\text{eff}}(\theta, \bar{\theta}; \mathbf{r}^{(1)},
    \mathbf{r}^{(2)}) 
    &\;=\; \mathcal{S}_{0}^{a}(\theta^{(1)}, \bar{\theta}^{(1)}) + 
    \mathcal{S}_{0}^{a}(\theta^{(2)}, \bar{\theta}^{(2)}) -\int_{x, x'} \, 
    \bar{\eta}(x) \, S_F^{(0)}(x - x') \,
    \eta(x') \, ,
\end{align}
where $S_F^{(0)}(x - x')$ is the free propagator of the Dirac field, and
we have introduced:
\begin{equation}\label{eq:FermSource}
    \eta(x)
    \;\equiv\;\lambda \, \big( \, \delta^{d}(\mathbf{x} -
    \mathbf{r}^{(1)}) \, \theta^{(1)}(\tau) + \delta^{d}(\mathbf{x} -
    \mathbf{r}^{(2)}) \, \theta^{(2)}(\tau) \, \big) \,, 
\end{equation}
and its Dirac adjoint.

Substituting (\ref{eq:FermSource}) into equation (\ref{eq:SeffAA}),
we observe that the intermediate effective action contains a diagonal
part in the atomic fields, $\mathcal{S}_{\text{eff}}^{\text{d}}$,
and a non-diagonal part, $\mathcal{S}_{\text{eff}}^{\text{nd}}$,
in which the fields are mixed:
\begin{equation}
    \mathcal{S}_{\text{eff}}
    \;=\;
    \mathcal{S}_{\text{eff}}^{\text{d}}
    \;+\; \mathcal{S}_{\text{eff}}^{\text{nd}} \; .
\end{equation}
The diagonal part takes the form:
\begin{align}
    \mathcal{S}_{\text{eff}}^{\text{d}}
    &\;=\;
    \sum_{\alpha} \, \int_{\tau} \bar{\theta}^{\,(\alpha)}(\tau) \,
    \big( \, \partial_\tau + \Omega \, \big) \, \theta^{(\alpha)}(\tau)
    \;+\;\big[\mathcal{S}_{\text{eff}}^{\text{d}}\big]_{\text{div}} \, .
\end{align}
The first term in this expression represents the sum of the free actions
of the atoms. The second term, arising from their self-energies, is a
divergent contribution that leads to a renormalization of the
frequency $\Omega$. Since this divergent term is
independent of the positions $\mathbf{r}^{(1)}$ and $\mathbf{r}^{(2)}$,
it can be discarded following a procedure analogous to that described in
\cite{paper:FoscoHansen2022}.

The non-diagonal part, which describes the coupling between the
fluctuations of the two atoms, is:
\begin{align}
    \mathcal{S}_{\text{eff}}^{\text{nd}}
    \;=\; -
    \lambda^2 \, \sum_{\alpha \neq \beta} \, \int_{\tau,\tau^\prime} \,
    \bar{\theta}^{\,(\alpha)}(\tau) \,
    S_F^{(0)}({\tau - \tau^\prime,\mathbf{r}^{(\alpha)}
    - \mathbf{r}^{(\beta)}}) \, \theta^{(\beta)}(\tau^\prime) \, .
\end{align}

By combining the diagonal and off-diagonal terms and performing a Fourier
transform to the frequency domain, we have:
\begin{align}
    \mathcal{S}_{\text{eff}}
    \;=\;
    \sum_{\alpha,\beta} \, \int \frac{d\nu}{2\pi} \;\;
    \widetilde{\hspace{-0.1em} \bar{\theta}}
    {\vphantom{\theta}}^{\,(\alpha)}(\nu) \,
    M^{(\alpha\beta)} (\nu,\mathbf{r}) \,
    \widetilde{\theta}^{\,(\beta)}(\nu) \,,
\end{align}
where the matrix $M^{(\alpha\beta)}(\nu, \mathbf{r})$ is defined as:
\begin{equation}
    M^{(\alpha\beta)}(\nu,\mathbf{r})
    \;\equiv\;
    \delta^{(\alpha\beta)} \, \widetilde{\Delta}^{-1}(\nu,\Omega)
    - \lambda^2 \, \sigma^{(\alpha\beta)} \, \widetilde{S}_{F}^{(0)}(\nu;
    \mathbf{r}^{(\alpha)} - \mathbf{r}^{(\beta)}) \, ,
\end{equation}
with $\sigma^{(\alpha\beta)} \equiv \delta^{(\alpha 1)}
\delta^{(\beta 2)} + \delta^{(\alpha 2)} \delta^{(\beta 1)}$,
$\mathbf{r} \equiv \mathbf{r}^{(1)} - \mathbf{r}^{(2)}$, 
and:
\begin{equation}
    \widetilde{\Delta}(\nu,\Omega)
    \;\equiv\;
    \frac{i\nu + \Omega}{\nu^2 + \Omega^2} \, .
\end{equation}

Finally, by integrating out the atomic fields, we obtain the full
effective action for the system:
\begin{align}
    \label{eq:EI_AA}
    e^{-\Gamma_{\text{eff}}(\mathbf{r}^{(1)},\mathbf{r}^{(2)})}
    \;=\;
    \det\,[\,M^{(\alpha\beta)}(\nu,\mathbf{r})\,] \, ,
\end{align}
where the determinant must be taken over frequencies, atomic indices,
and Dirac matrix indices.

\section{Interaction energy}\label{sec:vanderwaals}
Having obtained the effective action of the system, we can now compute
the interaction energy $E_I(\mathbf{r})$ between the atoms.
We find that interaction energy to be:
\begin{align}
    \label{eq:EI_vdw}
    E_I(\mathbf{r})
    \;=\;
    \lim_{T\to\infty} \frac{\Gamma_{\text{eff}}(\mathbf{r})}{T}
    \;=\;
    - \int_{-\infty}^{\infty} \frac{d\nu}{2\pi} \, \log \det \, 
    [\,\mathds{1} - \mathds{T}(\nu,\mathbf{r})\,] \, ,
\end{align}
where $T$ is the Euclidean time extent, and the matrix $\mathds{T}$
encodes the correlations between the quantum fluctuations of the atoms,
mediated by the Dirac field. This matrix is defined as:
\begin{equation}
    \label{eq:Tmatrix_AA}
    \mathds{T}(\nu,\mathbf{r})
    \;\equiv\;
    \lambda^4 \, \widetilde{\Delta}(\nu, \Omega) \, \widetilde{S}_{F}^{(0)}
    (\nu;\mathbf{r}) \, \widetilde{\Delta}(\nu, \Omega) \,
    \widetilde{S}_{F}^{(0)}
    (\nu;-\mathbf{r}) \, .
\end{equation}

The free fermion propagator $\widetilde{S}_F^{(0)}(\nu; \mathbf{r})$,
appearing in the expression for $\mathds{T}$, takes the general form:
\begin{equation}
    \widetilde{S}_{F}^{(0)}(\nu;\mathbf{r})
    \;=\;
    a(\nu,r) \, i \gamma_0 + \frac{1}{r} \, b(\nu,r) \, r_i \gamma_i \, ,
\end{equation}
valid for any dimension $d$. The functions $a(\nu,r)$ and $b(\nu,r)$
encapsulate the spatial dimensional dependence, with their explicit
forms provided in Appendix A.

Substituting the propagator into the determinant of (\ref{eq:EI_vdw}),
the resulting expression is:
\begin{align}
    d(\nu,r)
    &\;\equiv\;
    \det \left[ \mathds{1} - \mathds{T}(\nu,\mathbf{r})
    \right] \nonumber\\
    &\;=\; \left[
    1 + 2\,\lambda^4\,\widetilde{\Delta}^2(\nu,\Omega)\,\left(a^2(\nu,r)
    + b^2(\nu,r)\right) \right. \nonumber\\
    &\quad\quad\quad +
    \lambda^8\,\widetilde{\Delta}^4(\nu,\Omega)\,\left(a^2(\nu,r)
    - b^2(\nu,r)\right)^2 \bigg]^{\lfloor \frac{d+1}{2} \rfloor} \, ,
    \label{eq:d_nu_r}
\end{align}
where $\lfloor \cdot \rfloor$ denotes the floor function.

Since $d^{*}(\nu,r) = d(-\nu,r)$, the interaction energy simplifies to:
\begin{equation}
    E_I(r)
    \;=\;
    - \int_{0}^{\infty} \frac{d\nu}{2\pi} \,
    \log\big(|d(\nu,r)|^2\big) \, ,
\end{equation}
which is clearly real, indicating the stability of the two-atom system.
In contrast, in the model considered in \cite{paper:FoscoHansen2022},
where the atoms have internal structure, an imaginary component appears
in the interaction energy, signaling the instability of the system.
However, in the present case, since the atoms are modeled as point-like,
such instability cannot arise.

To simplify the expression further, we perform a change of variables,
$\nu = \frac{u}{r}$, and introduce the dimensionless parameter
$x \equiv \Omega r$, which denotes the interatomic distance in units
of the atomic transition wavelength $\sim \Omega^{-1}$.
Also, we rescale the coupling as
$\lambda = \widetilde{\lambda} \, \Omega^{1-d/2}$, thereby providing a
dimensionless description of the interaction strength.
For $d = 1$ and $d = 2$, the argument of the logarithm remains finite as
$x \to 0$, allowing for a straightforward expansion in powers of
$\widetilde{\lambda}^4$.
The interaction energy can then be expressed as
$E_I(r) = \Omega \, \mathcal{E}_I(\Omega r)$, where the dimensionless
interaction energy $\mathcal{E}_I(x)$ is given by:
\begin{equation}
    \label{eq:EIadim_AA}
    \mathcal{E}_I(x)
    = -\frac{1}{x} \int_{0}^{\infty} \frac{du}{2\pi} \,
    \log\big(|D(u,x)|^2\big) \, ,
\end{equation}
where:
\begin{align}
    D(u,x)
    \;\equiv\;
    \bigg[\,
    1 &+ 2\,\widetilde{\lambda}^4 \, x^{2(2-d)} \widetilde{\Delta}^2(u,x)
    \, \big(a^2(u,1) + b^2(u,1)\big) \nonumber \\
    &+ 4\,\widetilde{\lambda}^8 \, x^{4(2-d)} \widetilde{\Delta}^4(u,x)
    \, {\big(a^2(u,1) - b^2(u,1)\big)}^2 \, 
    \bigg]^{\lfloor \frac{d+1}{2} \rfloor}
     \, .
\end{align}

We will analyze this expression in detail under various assumptions
regarding the system's parameters.

\subsection{Weak coupling limit}
To further understand the behavior of the system, we analyze the interaction 
energy under the weak coupling limit. This regime simplifies the equations, 
offering insight into both short- and long-distance behaviors.

In the weak coupling regime, characterized by
$\widetilde{\lambda}^4 \ll 1$, the interaction energy simplifies to a
logarithmic form, $\log(1 + z)$, where the terms contributing to $z$ are
proportional to powers of $\widetilde{\lambda}$.

At large distances ($x \gg 1$), the interaction energy becomes
exponentially suppressed, indicating a weakening of the interaction as
the separation between the atoms increases. In the short-distance limit
($x \to 0$), the behavior of the interaction varies with the spatial
dimension $d$. For $d = 1$ and $d = 2$, the argument of the logarithm 
remains finite as $x \to 0$, allowing for a straightforward expansion in
powers of $\widetilde{\lambda}^4$. However, in $d = 3$, the argument of the 
logarithm develops divergences as $x \to 0$, requiring the additional
condition $\widetilde{\lambda}^4/x^2 \ll 1$ to ensure the validity of the
expansion and to avoid such divergences.

In this regime, the logarithmic term $\log(1 + z)$ can be approximated
by $z$ when $|z| \ll 1$, further simplifying the interaction energy.
Applying this approximation, the interaction energy reduces to:
\begin{equation}
    \label{eq:EIadim_AAweak}
    \mathcal{E}_I(x) \sim - \widetilde{\lambda}^4 \, x^{3-2d} \,
    2^{1 + \lfloor \frac{d+1}{2} \rfloor} \int_{-\infty}^{\infty}
    \frac{du}{2\pi} \, \frac{\big(x^2 - u^2\big) \, \big(a^2(u,1)
    + b^2(u,1)\big)}{\big(u^2 + x^2\big)^2} \, .
\end{equation}

Table \ref{table:EIAAweak} presents the explicit results of the integral
for different spatial dimensions $d$, illustrating how the dimensionless
interaction energy $\mathcal{E}_I(x)$ depends on the interatomic distance
$x$ in each case.
\begin{table}[!ht]
\centering
\setlength{\tabcolsep}{15pt}
\begin{tabular}{cc}
\toprule
\midrule
\noalign{\vskip 2pt}
$d$ & $\mathcal{E}_I(x)$ \\
\noalign{\vskip 2pt} 
\midrule
\noalign{\vskip 4pt}
1 & $- \frac{\widetilde{\lambda}^4}{\pi} \, 2x \, g(2x)$  \\
\noalign{\vskip 6pt}
2 & $\begin{aligned}
    {\textstyle  -\frac{\widetilde{\lambda}^4}{2 \pi x} \,
    \frac{\sqrt{\pi}}{4 x \, \pi^2}} \,
    \big\{
    \, x^2 \, &\big[\,
    {G^{2, 0}_{0, 3}} \left( x^2 \middle|
    \scalebox{0.7}{$\begin{array}{c} -\frac{1}{2} \\ 0, 0, 0 \end{array}$}
    \right)
    \;+\;
    x^2 \,
    {G^{2, 0}_{1, 4}} \left( x^2 \middle|
    \scalebox{0.7}{$\begin{array}{c} -\frac{3}{2}, \frac{1}{2} \\
    -\frac{1}{2}, 0, 0, 0 \end{array}$}
    \right) \big] \nonumber\\
    -\, &\big[\,
    {G^{2, 0}_{0, 3}} \left( x^2 \middle|
    \scalebox{0.7}{$\begin{array}{c} -\frac{1}{2} \\ -1, 0, 1
    \end{array}$} \right)
    \;+\;
    x^2 \,
    {G^{2, 0}_{1, 4}} \left( x^2 \middle|
    \scalebox{0.7}{$\begin{array}{c} -\frac{3}{2}, \frac{1}{2} \\
    -1, -\frac{1}{2}, 0, 1 \end{array}$}
    \right) \big] \,
    \big\}
    \end{aligned}$ \\
\noalign{\vskip 6pt}
3 & $- \frac{\widetilde{\lambda}^4}{4\pi^3 x^3} \, \big(1 - (2x)^2 \,
g(2x)\big)$ \\
\noalign{\vskip 4pt}
\midrule
\bottomrule
\end{tabular}
\caption{Dimensionless interaction energy $\mathcal{E}_I(x)$ between the
two fermionic atoms in the weak coupling regime, as a function of the
dimensionless distance $x$ and for different spatial dimensions $d$.
In this context, we use the auxiliary function
$g(z) \equiv -\text{Ci}(z) \, \cos(z) - \text{si}(z) \, \sin(z)$, where
$\text{si}(z) \equiv \text{Si}(z) - \frac{\pi}{2}$, with $\text{Ci}(z)$
and $\text{Si}(z)$ being the cosine and sine integrals,
respectively \cite{book:AbramowitzStegun1964}. Additionally, $G$
corresponds to the Meijer G-function
\cite{book:Andrews1985,book:Bateman1953}.}
\label{table:EIAAweak}
\end{table}

Based on the results and the behavior of the functions presented in
Table~\ref{table:EIAAweak}, we conclude that the interaction energy is
negative for all spatial dimensions $d$ and at any distance $x$.
Moreover, the magnitude of the energy decreases monotonically as the
distance $x$ increases, indicating that the force between the atoms is
consistently attractive across all regimes considered. This attractive
force is stronger at short distances and gradually weakens as the atoms
move farther apart, in line with the expected behavior in the weak
coupling regime.

We now proceed to analyze the behavior of the interaction energy in both
the long-distance and short-distance limits, providing insights into how
the energy scales as a function of the interatomic separation in these
two regimes.

\subsubsection{Long-distance limit ($x \gg 1$)}
When the separation between the atoms is much larger than the
characteristic wavelength of atomic transitions, $\sim \Omega^{-1}$,
i.e., $r \gg \Omega^{-1}$, retardation effects become significant due to
the finite propagation speed of the interaction. As the distance between
the atoms increases, the interaction energy decays more rapidly, with the
asymptotic behavior dominated by the leading-order terms in $1/x$.

To capture this behavior, we analyze the asymptotic form of the functions
appearing in the expression for the interaction energy, as summarized in
Table~\ref{table:EIAAweak}. The results for various spatial dimensions
are shown in Table~\ref{table:EIAAlimit}, which highlights the power-law
dependence of the interaction energy at large distances.

\subsubsection{Short-distance limit ($x \ll 1$)}

At very short distances, $r \ll \Omega^{-1}$, the separation between the
atoms is much smaller than the characteristic wavelength of atomic
transitions, and retardation effects become negligible. In this regime,
known as the \textit{London limit}~\cite{book:Milonni1994}, the
interaction can be treated as essentially instantaneous.

To evaluate the interaction energy, we approximate $x \simeq 0$ in the
numerator of the integrand in (\ref{eq:EIadim_AAweak}), which is valid
due to the small atomic separation.

The results are summarized in Table~\ref{table:EIAAlimit}, where the
power-law dependence of the interaction energy at short distances shows
a slower decay compared to the long-distance limit.

\begin{table}[h!]
    \centering
    \setlength{\tabcolsep}{15pt}
    \begin{tabular}{ccc}
    \toprule
    \midrule
    \noalign{\vskip 3pt}
    \multicolumn{3}{c}{\small $\mathcal{E}_I(x)$} \\
    \noalign{\vskip 3pt}
    \midrule
    \noalign{\vskip 2pt}
    $d$ & $x \gg 1$ & $x \ll 1$ \\
    \noalign{\vskip 2pt}
    \midrule
    \noalign{\vskip 2pt}
    1 & $- \frac{1}{2\pi} \, \frac{\widetilde{\lambda}^4}{x}$ &
    $-\frac{1}{4} \, \widetilde{\lambda}^4$ \\
    \noalign{\vskip 6pt}
    2 & $- \frac{1}{16\pi} \, \frac{\widetilde{\lambda}^4}{x^3}$ &
    $- \frac{1}{8\pi^2} \, \frac{\widetilde{\lambda}^4}{x^2}$\\
    \noalign{\vskip 6pt}
    3 & $- \frac{24}{(4\pi)^3} \, \frac{\widetilde{\lambda}^4}{x^5}$ &
    $- \frac{1}{(4\pi)^2} \, \frac{\widetilde{\lambda}^4}{x^4}$ \\
    \noalign{\vskip 2pt}
    \midrule
    \bottomrule
    \end{tabular}
    \caption{Dimensionless interaction energy $\mathcal{E}_I(x)$ between
    the fermionic atoms in the weak coupling regime, as a function of the
    dimensionless distance $x$, for different spatial dimensions $d$, in
    the short-distance ($x \ll 1$) and long-distance ($x \gg 1$) limits.}
    \label{table:EIAAlimit}
\end{table}

\section{Atom-mirror interaction}\label{sec:casimir-polder}

Let us study here the interaction between an atom and a perfectly 
reflecting boundary, modeled using bag-model boundary conditions. 
The wall is located at $x_d = 0$, what is obtained by equipping  
the Dirac field $\psi$ with the action:
\begin{equation}
    \mathcal{S}_0^D(\psi, \bar{\psi})
    \;=\;
    \int d^{d+1}x \, \bar{\psi}(x) \, \big( \, \slashed{\partial}
    + V(x) \, \big) \, \psi(x) \,.
\end{equation}
Here, the presence of $V(x) = g \, \delta(x_d)$ enforces bag model boundary 
conditions when $g = 2$. For any other value of $g$, the boundary
conditions become imperfect, leading to partial transmission of the
fermionic current through the wall
\cite{paper:FoscoLosada2008,paper:FoscoHansen2023}.

The interaction between the two-level system and the Dirac field occurs
at a fixed position $\mathbf{r} = (0, \ldots, a) \in \mathds{R}^{d}$,
with $a$ being the separation between the atom and the reflecting
surface. This interaction is described by:
\begin{equation}
    \mathcal{S}_{I}
    \;=\;
    \lambda \int d\tau \left( \, \bar{\psi}(\tau, \mathbf{r})
    \, \theta(\tau)
    + \bar{\theta}(\tau) \, \psi(\tau, \mathbf{r}) \, \right) \,,
\end{equation}
where $\lambda$ determines the strength of the coupling between the
atom and the fermionic field.

The total action for the system, which includes the two-level atom,
the Dirac field with the boundary condition, and their interaction,
is given by:
\begin{equation}
    \mathcal{S}
    \;=\;
    \mathcal{S}_0^a(\theta, \bar{\theta})
    \;+\; \mathcal{S}_0^D(\psi, \bar{\psi})
    \;+\; \mathcal{S}_{I}(\theta, \bar{\theta}, \psi, \bar{\psi}; a) \,.
\end{equation}

Now, we will examine the effective description of the system alongside
the interaction energy.

\subsection{Effective Action and Interaction Energy}

To compute the interaction energy between the atom and the mirror, we
first derive the effective action by integrating out the Dirac field.
The effective action depends on the distance $a$ between the atom and
the mirror and is given by:
\begin{align}
    e^{-\Gamma_{\text{eff}}(a)}
    &\;\equiv\;
    \frac{1}{\mathcal{N}} \, \mathcal{Z}(a) \, , \\
    \mathcal{Z}(a)
    &\;=\;
    \int \mathcal{D}\theta \mathcal{D}\bar{\theta} \mathcal{D}\psi
    \mathcal{D}\bar{\psi} \,
    e^{-\mathcal{S}(\theta,\bar{\theta},\psi,\bar{\psi};a)} \, .
\end{align}

By performing the functional integral over the Dirac field, we obtain the
intermediate effective action:
\begin{align}
    \mathcal{S}_{\text{eff}}(\theta,\bar{\theta};a)
    &\;=\;
    \mathcal{S}_{0}^{a}(\theta,\bar{\theta}) \nonumber \\
    &\;-\;
    \lambda^2 \, \lim_{\mathbf{x}_{\shortparallel}^{\prime}
    \to \, \mathbf{x}_{\shortparallel}^{\vphantom{\prime}}} \,
    \int_{\tau,\tau^\prime} \bar{\theta}(\tau) \,
    S_F(\tau-\tau^\prime,\mathbf{x}_{\shortparallel}^{\vphantom{\prime}}
    - \mathbf{x}_{\shortparallel}^{\prime};a,a) \,
    \theta(\tau^\prime) \, ,
\end{align}
where $S_F(\tau-\tau^\prime,\mathbf{x}_{\shortparallel}^{\vphantom{\prime}}
-\mathbf{x}_{\shortparallel}^{\prime};x_d,x_d^{\prime})$ is the propagator
of the Dirac field in the presence of the mirror. This propagator can be
decomposed as:
\begin{equation}
    S_F(x;x^\prime)
    \;=\;
    S_F^{(0)}(x-x^\prime) \;+\; T_F(x;x^\prime) \,,
\end{equation}
where $S_F^{(0)}(x-x^\prime)$ is the free propagator, and
$T_F(x;x^\prime)$ accounts for the mirror's contribution.
A detailed derivation of this propagator can be found in Appendix B.

As in the atom-atom interaction case, the free propagator
$S_F^{(0)}(x-x^\prime)$ introduces a divergent term related to the
atom's self-energy. This divergence, which occurs when
$\mathbf{x}_{\shortparallel}^{\prime} \to
\mathbf{x}_{\shortparallel}^{\vphantom{\prime}}$, leads to a
renormalization of the energy $\Omega$, and since it is independent of
the atom's position relative to the mirror, we discard it using the same
procedure as before.

The relevant term for computing the interaction energy is the
mirror-dependent contribution
$T_F(\tau-\tau^\prime,\mathbf{0}_{\shortparallel};a,a)$, which
explicitly depends on the distance $a$ and takes the form:
\begin{align}
    T_F(\tau-\tau^\prime,\mathbf{0}_{\shortparallel};a,a)
    \;=\;
    S_{F}^{(0)}
    (\tau - \tau^\prime,\mathbf{0}_{\shortparallel},2a) \, \gamma_d \, .
\end{align}

To simplify further calculations, we transform the effective action to
the frequency domain via a Fourier transform, resulting in:
\begin{equation}
    \mathcal{S}_{\text{eff}}(\widetilde{\theta},
    \widetilde{\hspace{-0.1em} \bar{\theta}};a)
    \;=\;
    \int \frac{d\nu}{2\pi} \,\,
    \widetilde{\hspace{-0.1em} \bar{\theta}}(\nu) \,
    \big( \, \widetilde{\Delta}^{-1}(\nu,\Omega) \,
    - \lambda^2 \, \widetilde{T}_{F}(\nu;
    \mathbf{0}_{\shortparallel};a,a) \, \big) \,
    \widetilde{\theta}(\nu) \, ,
\end{equation}
where $\widetilde{T}_{F}(\nu; \mathbf{0}_{\shortparallel};a,a)$ is the
Fourier transform of the mirror's contribution.

Finally, we obtain the full effective action by integrating over the
atomic degrees of freedom:
\begin{align}
    e^{-\Gamma_{\text{eff}}(a)}
    \;=\;
    \frac{1}{\mathcal{N}} \int \mathcal{D}\widetilde{\theta}
    \mathcal{D} \kern 0.1em \widetilde{\hspace{-0.1em} \bar{\theta}} \,
    e^{-\mathcal{S}_{\text{eff}}
    (\widetilde{\theta}, \kern 0.1em
    \widetilde{\hspace{-0.1em}\bar{\theta}};a)}
    \;=\;
    \det \big[ \,
    \mathds{1} - \mathds{T}(\nu,a)
    \, \big] \, ,
\end{align}
where the matrix $\mathds{T}(\nu,a)$ is defined as:
\begin{align}
    \label{eq:Tmatriz_AP}
    \mathds{T}(\nu,a)
    \;\equiv\;
    \lambda^2 \, \widetilde{\Delta}(\nu,\Omega) \,
    \widetilde{S}_{F}^{(0)}(\nu;\mathbf{0}_{\shortparallel},2a)
    \, \gamma_d \, .
\end{align}

The interaction energy can then be expressed similarly to equation
(\ref{eq:EI_AA}), but with the matrix $\mathds{T}$ specific to this
scenario. Using dimensionless quantities, we obtain a result analogous
to equation (\ref{eq:EIadim_AA}), with the function $D(u,x)$ defined as:
\begin{align}
    D(u,x)
    \;\equiv\;
    \bigg[\, 1 
    &- 2 \, \widetilde{\lambda}^2 \, x^{2-d}\, \widetilde{\Delta}(u,x)
    \, b(u,2) \nonumber\\
    &- \widetilde{\lambda}^4 \, x^{2(2-d)} \,
    \widetilde{\Delta}^2(u,x)\,\big(a^2(u,2) - b^2(u,2)\big)^2 \,
    \bigg]^{\lfloor \frac{d+1}{2} \rfloor} \, .
\end{align}

The interaction energy is, once again, real for all distances $x$,
confirming the stability of the system across the entire range of
distances and coupling strengths.

\subsection{Weak coupling limit}
In the weak coupling regime, the analysis is analogous to the case of the
two-atom interaction discussed in Sect.~\ref{sec:vanderwaals}. In this
case, the expression for the interaction energy is:
\begin{equation}
    \mathcal{E}_{I}(x)
    \;\sim\;
    2^{1 + \lfloor \frac{d+1}{2} \rfloor} \, \widetilde{\lambda}^2 \,
    x^{2-d} \, \int_{-\infty}^{\infty} \frac{du}{2\pi} \,
    \frac{b(u,2)}{u^2 + x^2} \, .
\end{equation}
From this result, we observe that, unlike the two-atom case, the
interaction energy is positive. Since the energy monotonically decreases
with distance $x$, this indicates the presence of a repulsive force
between the atom and the conducting plane.

In Table \ref{table:EIAP}, we present the expressions obtained for
the interaction energy in the weak coupling limit, and in Table
\ref{table:EIAPlimits}, we show the results in the long- and
short-distance limits.

\begin{table}[H]
    \centering
    \setlength{\tabcolsep}{15pt}
    \begin{tabular}{cc}
    \toprule
    \midrule
    \noalign{\vskip 2pt}
    $d$ & $\mathcal{E}_I(x)$ \\
    \noalign{\vskip 2pt}
    \midrule
    \noalign{\vskip 4pt}
    1 & $\frac{1}{\pi} \, \widetilde{\lambda}^2 \, f(2x)$  \\
    \noalign{\vskip 6pt}
    2 & $\frac{1}{4} \, \widetilde{\lambda}^2 \,
        \big[\, Y_{-1}(2x) - \mathbf{H}_{-1}(2x) \,\big]$ \\
    \noalign{\vskip 6pt}
    3 & $\frac{1}{(2 \pi x)^2} \, \widetilde{\lambda}^2
        \, \big[ \, f(2x) + 2x \, g(2x) \,\big]$ \\
    \noalign{\vskip 4pt}
    \midrule
    \bottomrule
\end{tabular}
\caption{Dimensionless interaction energy $\mathcal{E}_I(x)$ between
the plane and the atom in the weak coupling regime, as a function of
the dimensionless distance $x$ and for different spatial dimensions $d$.
In this context, we use the auxiliary function
$f(z) \equiv \text{Ci}(z) \, \sin(z) - \text{si}(z) \, \cos(z)$, along
with the previously defined function $g(z)$. Additionally, $Y_{\nu}(z)$
corresponds to the Bessel function of the second kind, and
$\mathbf{H}_{\nu}(z)$ is the Struve function
\cite{book:AbramowitzStegun1964}.}

\label{table:EIAP}
\end{table}

\begin{table}[H]
    \centering
    \setlength{\tabcolsep}{15pt}
    \begin{tabular}{ccc}
    \toprule
    \midrule
    \noalign{\vskip 2pt}
    \multicolumn{3}{c}{$\mathcal{E}_I(x)$} \\
    \noalign{\vskip 2pt}
    \midrule
    \noalign{\vskip 4pt}
    $d$ & $x \gg 1$ & $x \ll 1$ \\
    \midrule
    \noalign{\vskip 6pt}
    1 & $\frac{1}{2\pi} \, \frac{\widetilde{\lambda}^2}{x}$ &
    $\frac{1}{2} \, \widetilde{\lambda}^2$ \\
    \noalign{\vskip 6pt}
    2 & $\frac{1}{8\pi} \, \frac{\widetilde{\lambda}^2}{x^2}$ &
    $\frac{1}{4\pi} \, \frac{\widetilde{\lambda}^2}{x}$ \\
    \noalign{\vskip 6pt}
    3 & $\frac{1}{(2\pi)^2} \, \frac{\widetilde{\lambda}^2}{x^3}$ &
    $\frac{1}{4\pi} \, \frac{\widetilde{\lambda}^2}{x^2}$ \\
    \noalign{\vskip 4pt}
    \midrule
    \bottomrule
\end{tabular}
\caption{Dimensionless interaction energy $\mathcal{E}_I(x)$ between the
fermionic atom and the perfectly reflecting plane in the weak coupling
regime, as a function of the dimensionless distance $x$, for different
spatial dimensions $d$, in the short-distance ($x \ll 1$) and
long-distance ($x \gg 1$) limits.}
\label{table:EIAPlimits}
\end{table}

\section{Conclusions}\label{sec:conclusions}

In this work, we have introduced and studied the fermionic counterparts
of the van der Waals and Casimir-Polder interactions in two distinct 
setups. First, we analyzed the interaction energy between two fermionic
two-level atoms mediated by a massless fermionic field, deriving integral
expressions for the interaction energy in 1, 2, and 3 spatial dimensions.
Our results show that the interaction is universally attractive, becoming
stronger at short distances and decaying more rapidly at large distances,
in line with the expected power-law behavior.

We then examined the interaction between a fermionic two-level atom and
a perfectly reflecting fermionic mirror, modeled using bag boundary
conditions. In contrast to the two-atom case, this interaction is
repulsive. We again derived integral expressions for the interaction 
energy in various dimensions, observing that the repulsive force dominates
at short distances and weakens as the separation increases, following a
characteristic decay.

Our findings underscore the relevance of fermionic field vacuum fluctuations in 
quantum interactions. These could pave the way for exploring analogous 
phenomena in experimental setups, such as in graphene or other condensed matter 
systems.


\section*{Acknowledgements}
The authors thank ANPCyT, CONICET and UNCuyo for financial support.

\section*{Appendix A: Free Dirac propagator}\label{sec:appendix_A}

Here, we present the Euclidean propagator for a free Dirac field.
This propagator satisfies the following differential equation:
\begin{equation}
    (\,\slashed{\partial}\,)\,S_{F}^{(0)}(x;x^\prime)
    \;=\;
    \delta^{d+1}(x-x^\prime) \, . 
\end{equation}

Owing to the invariance of the system under spacetime translations, the
propagator can be expressed in terms of its Fourier transform:
\begin{align}
    S_{F}^{(0)}(x-x^\prime)
    \;=\;
    \int \frac{d\nu}{2\pi} \int \frac{d^{d}\mathbf{k}}{(2\pi)^{d}} \,
    \widetilde{S}_{F}^{(0)}(\nu; \mathbf{r}) \, ,
\end{align}
where, in an arbitrary spatial dimension $d$, $\widetilde{S}_{F}^{(0)}$
has the form:
\begin{equation}
    \widetilde{S}_F^{(0)}(\nu;\mathbf{r})
    \;=\;
    a(\nu,r) \, i \gamma_0 + \frac{1}{r} \, b(\nu,r) \, r_i \gamma_i \, ,
\end{equation}
with the functions $a(\nu,r)$ and $b(\nu,r)$ depending explicitly on $d$.
The explicit forms of these functions for $d = 1, 2, 3$ are provided in
Table~\ref{table:free_propagator}.

\begin{table}[H]
    \centering
    \setlength{\tabcolsep}{15pt}
    \begin{tabular}{ccc}
    \toprule
    \midrule
    \noalign{\vskip 3pt}
    $d$ & $a(\nu,r)$ & $b(\nu,r)$ \\
    \noalign{\vskip 3pt}
    \midrule
    \noalign{\vskip 2pt}
    1 & $\frac{1}{2} \, \text{sign}(\nu) \, e^{-r|\nu|}$ & $\frac{1}{2}
    \, e^{-r|\nu|}$ \\
    \noalign{\vskip 6pt}
    2 & $\frac{\nu}{2} \, K_0(r|\nu|)$ & $\frac{\nu}{2} \, K_1(r|\nu|)$ \\
    \noalign{\vskip 6pt}
    3 & $\frac{\nu}{4\pi r} \, e^{-r|\nu|}$ & $\frac{1 + r|\nu|}{4\pi r^2}
    \, e^{-r|\nu|}$ \\
    \noalign{\vskip 2pt}
    \midrule
    \bottomrule
    \end{tabular}
    \caption{Functions $a(\nu, r)$ and $b(\nu, r)$ for different spatial
    dimensions $d$. Here, $K_\nu$ denotes the modified Bessel function of
    the second kind.}
    \label{table:free_propagator}
\end{table}

\section*{Appendix B: Dirac propagator in the presence
of a Wall}\label{sec:appendix_B}

In this appendix, we derive the Euclidean Dirac propagator in the presence
of a wall located at $x_d = 0$. With this setup, the propagator satisfies
the following differential equation:
\begin{equation}
    \big( \, \slashed{\partial} + g\,\delta(x_d) \, \big) \,
    S_{F}(x;x^\prime) \;=\; \delta^{d+1}(x-x^\prime) \, .
\end{equation}

Given that the system is time-invariant and translation-invariant along
the directions parallel to the wall, $\mathbf{x}_{\shortparallel}$, the
propagator can be expressed in terms of its Fourier transform along
those coordinates:
\begin{align}
    S_{F}(x;x^\prime)
    &\;=\;
    S_{F}(\tau-\tau^\prime, \mathbf{x}_{\shortparallel}
    - \mathbf{x}_{\shortparallel}^{\prime}; x_d, x_d^{\prime})
    \nonumber \\
    &\;=\;
    \int \frac{d\nu}{2\pi} \int \frac{d^{d-1}
    \mathbf{k}_{\shortparallel}}{(2\pi)^{d-1}}
    e^{-i \nu(\tau-\tau^\prime) - i\mathbf{k}_{\shortparallel}
    (\mathbf{x}_{\shortparallel} - \mathbf{x}_{\shortparallel}^{\prime})}
    \widetilde{S}_F(\nu,\mathbf{k}_{\shortparallel};x_d,x_d^{\prime}) \, .
\end{align}
In Fourier space, the propagator satisfies:
\begin{equation}
    \big( -i \slashed{\nu} - i\slashed{\mathbf{k}}_{\shortparallel}
    + \slashed{\partial}_{x_d} + g \, \delta(x^d) \, \big) \,
    \widetilde{S}_{F}(\nu,\mathbf{k}_{\shortparallel};x_d,x_d^{\prime})
    \;=\;
    \delta(x_d - x_d^{\prime}) \, .
\end{equation}

To solve this equation, we apply the free propagator
$\widetilde{S}_{F}^{(0)}$ to both sides from the left:
\begin{equation}
    \label{eq:SFtilde}
    \widetilde{S}_{F}(\nu,\mathbf{k}_{\shortparallel};x_d,x_d^{\prime})
    = \widetilde{S}_{F}^{(0)}
    (\nu,\mathbf{k}_{\shortparallel};x_d,x_d^{\prime})
    - g\,\widetilde{S}_{F}^{(0)}(\nu,\mathbf{k}_{\shortparallel};x_d,0) \,
    \widetilde{S}_{F}(\nu,\mathbf{k}_{\shortparallel};0,x_d^{\prime}) \, .
\end{equation}

We observe that (\ref{eq:SFtilde}) is an implicit equation for
$\widetilde{S}_F$, as it depends on
$\widetilde{S}_F(\nu, \mathbf{k}_{\shortparallel};0, x'_d)$.
By evaluating at $x_d = 0$, we can explicitly solve for
$\widetilde{S}_F(\nu, \mathbf{k}_{\shortparallel};0, x'_d)$, yielding:
\begin{equation}
    \widetilde{S}_{F}(\nu,\mathbf{k}_{\shortparallel};0,x_d^{\prime})
    \;=\; \big(1 + g \, \widetilde{S}_{F}^{(0)}
    (\nu,\mathbf{k}_{\shortparallel};0,0)\big)^{-1} \,
    \widetilde{S}_{F}^{(0)}(\nu,
    \mathbf{k}_{\shortparallel};0,x_d^{\prime}) \, .
\end{equation}

Substituting this result into (\ref{eq:SFtilde}), we arrive at:
\begin{align}
    \widetilde{S}_{F}(\nu,\mathbf{k}_{\shortparallel};x_d,x_d^{\prime})
    &\;=\;
    \widetilde{S}_{F}^{(0)}
    (\nu,\mathbf{k}_{\shortparallel};x_d,x_d^{\prime})
    + \widetilde{T}_{F}(\nu,\mathbf{k}_{\shortparallel};x_d,x_d^{\prime})
    \, , \\
    \widetilde{T}_{F}(\nu,\mathbf{k}_{\shortparallel};x_d,x_d^{\prime})
    &\;=\;
    - \, g \, \widetilde{S}_{F}^{(0)}(\nu,\mathbf{k}_{\shortparallel};x_d,0)
    \, \big[ \, \mathds{1} + g \, \widetilde{S}_{F}
    (\nu,\mathbf{k}_{\shortparallel};0,0) \, \big]^{-1} \nonumber \\
    &\quad\quad \widetilde{S}_{F}^{(0)}
    (\nu,\mathbf{k}_{\shortparallel};0,x_d^{\prime}) \, ,
\end{align}
where the inverse of the term in brackets is given by:
\begin{align}
    \big[ \, \mathds{1} + g \, \widetilde{S}_{F}
    (\nu,\mathbf{k}_{\shortparallel};0,0) \, \big]^{-1}
    \;=\;
    \frac{1}{1 + \left(\frac{g}{2}\right)^2}
    \left[
    \mathds{1} - i \, \frac{g}{2} \, \frac{\slashed{\nu}
    + \slashed{\mathbf{k}}_{\shortparallel}}{\sqrt{\nu^2
    + \mathbf{k}_{\shortparallel}^{2}}}
    \right] \, .
\end{align}

Finally, for $x_d, x_d^{\prime} > 0$, which corresponds to the region
where the atom is located, we find:
\begin{equation}
    \widetilde{T}_{F}(\nu,\mathbf{k}_{\shortparallel};x_d,x_d^{\prime})
    \;=\;
    \frac{g}{1 + {\left(\frac{g}{2}\right)}^2}
    \, \widetilde{S}_{F}^{(0)}
    (\nu,\mathbf{k}_{\shortparallel};x_d,-x_d^{\prime}) \, \gamma_d \, .
\end{equation}

In the case where the mirror perfectly reflects the normal component of
the fermionic current, corresponding to setting $g=2$, the expression
simplifies to:
\begin{equation}
    \widetilde{T}_{F}(\nu,\mathbf{k}_{\shortparallel};x_d,x_d^{\prime})
    \;=\;
    \widetilde{S}_{F}^{(0)}
    (\nu,\mathbf{k}_{\shortparallel};x_d,-x_d^{\prime}) \, \gamma_d \, .
\end{equation}


\end{document}